\documentstyle[vsolj01,graphicx]{article}

\begin{document}

\title{Outburst of Possible Eclipsing Symbiotic Variable AS 289}

\author{Kesao Takamizawa}
\author{(65-1 Oohinata, Saku-machi, Nagano 384-0502, Japan)}
\email{E-mail: GHA07243@nifty.ne.jp}

\author{Minoru Wakuda}
\author{(345 Kwabukuro, Ryuyo-cho, Iwata-gun, Sizuoka 438-0232 Japan)}
\email{E-mail: GZJ02446@nifty.ne.jp}

\author{Taichi Kato}
\author{(Dept. of Astronomy, Kyoto University, Kyoto 606-8502, Japan)}
\email{E-mail: tkato@kusastro.kyoto-u.ac.jp}

\section{Introduction}
   Takamizawa discovered an unusual variable star TmzV17 in the course of
his photographic patrol of variable stars (Takamizawa 1997).  The variable
(= GSC 5684.522, J2000.0 coordinates: 18\h 12\m 22\fs 13
-11\deg 40\arcm 07\farcs 1)
was later found to be identical with an emission line object AS 289
(Merrill and Burwell 1950).  AS 289 has been known as a symbiotic star
(Sanduleak and Stephenson 1973; Allen 1978) showing moderately strong
He II line and TiO absorption.  The object was also studied
spectrophotometrically by Blair et al. (1983).  Despite its relatively
abundant history of spectroscopy, neither outbursts nor variablity has been
reported.  A typical example of photometry of AS 289 is {\it V}=13.62 and
{\it B-V}=2.10 (Munari et al. 1992b).  The He II to H$\beta$ line ratio
suggests a hot ($\sim 1.5\times 10^5$ K) ionizing source (Kenyon 1986).

\section{Observations and results}
   The observations by Takamizawa were done at Saku Observatory
(Saku-machi, Nagano, Japan), using twin 10-cm patrol cameras (PENTAX 100SDUF,
fl=400mm) and T-Max400 films.  Wakuda's observations were done 
at his home, using a 10-cm camera (fl=400mm), T-Max and Tri-X films.
The both observers use consistently GSC stars for determining the magnitudes.

% Table 1
\begin{table}
\begin{center}
Table 1. Observations of AS 289 \\
\vspace{6pt}
\begin{tabular}{c c c c c c c c c}
\hline
JD-2400000 & mag & code$^*$ & JD-2400000 & mag & code$^*$ & JD-2400000 & 
mag & code$^*$ \\
\hline
45944.042 & 12.8 & Wdm & 49860.126 & 13.1 & Tmz & 50328.928 & 11.8 & Tmz \\
47267.272 & 14.5 & Wdm & 49860.134 & 12.5 & Wdm & 50372.899 & 12.0 & Tmz \\
47274.236 & 14.3 & Wdm & 49870.143 & 12.6 & Tmz & 50386.889 & 11.8 & Tmz \\
47961.317 & 14.3 & Wdm & 49888.047 & 12.4 & Tmz & 50494.342 & 12.6 & Tmz \\
47986.271 & 14.4 & Wdm & 49922.083 & 12.8 & Tmz & 50515.313 & 12.7 & Tmz \\
49485.182 & 14.9 & Tmz & 49934.081 & 12.9 & Tmz & 50578.190 & 12.8 & Tmz \\
49507.197 & 14.7 & Tmz & 49943.950 & 12.1 & Tmz & 50597.097 & 12.9 & Tmz \\
49520.219 & 14.3 & Wdm & 49950.062 & 11.9 & Tmz & 50606.201 & 13.0 & Tmz \\
49531.051 & 14.2 & Wdm & 49957.058 & 12.0 & Tmz & 50613.155 & 13.5 & Tmz \\
49537.046 & 14.3 & Tmz & 49972.945 & 12.1 & Tmz & 50629.128 & 13.2 & Tmz \\
49538.192 & 14.2 & Wdm & 49981.981 & 11.8 & Tmz & 50629.153 & 12.6 & Wdm \\
49546.119 & 14.2 & Tmz & 50002.912 & 12.1 & Tmz & 50633.144 & 12.9 & Tmz \\
49574.143 & 14.2 & Wdm & 50015.905 & 12.2 & Tmz & 50636.110 & 13.1 & Tmz \\
49575.041 & 14.7 & Tmz & 50110.340 & 13.0 & Tmz & 50644.178 & 13.1 & Tmz \\
49620.989 & 14.8 & Tmz & 50125.344 & 12.8 & Tmz & 50654.019 & 13.1 & Tmz \\
49648.892 & 14.0 & Tmz & 50133.314 & 12.6 & Tmz & 50682.985 & 13.2 & Tmz \\
49756.319 & 14.8 & Tmz & 50161.260 & 14.0 & Tmz & 50719.933 & 13.1 & Tmz \\
49783.273 & 14.8 & Tmz & 50186.199 & 13.9 & Tmz & 50740.914 & 13.0 & Tmz \\
49784.261 & 14.4 & Wdm & 50197.240 & 13.8 & Tmz & 50771.888 & 13.1 & Tmz \\
49788.284 & 14.9 & Tmz & 50229.191 & 12.6 & Tmz & 50875.299 & 12.7 & Tmz \\
49809.244 & 14.8 & Tmz & 50254.124 & 12.1 & Tmz & 50880.289 & 12.4 & Tmz \\
49815.243 & 14.8 & Tmz & 50277.101 & 11.9 & Tmz & 50907.283 & 12.2 & Wdm \\
49831.197 & 14.7 & Tmz & 50300.035 & 11.7 & Tmz & 50938.231 & 12.3 & Wdm \\
49834.242 & 13.7 & Tmz & 50310.983 & 11.9 & Tmz & & \\
\hline
\end{tabular}
\vskip 6pt
$^*$ code: Tmz (Takamizawa), Wdm (Wakuda)
\end{center}
\end{table}

   The resultant magnitude estimates are given in Table 1, the overall light
curve in Figure 1.  The light curve shows a sharp rise in brightness around
JD 2449834 (1985 April 26).  The most striking feature is a temporal fading
by more than two magnitudes between 1996 January and May.  Assuming the
symmetry, the minimum of the fading was determined as JD 2450176 $\pm$ 3
or 1996 April 2.  After the recovery from this episode, the system showed
a gradual fade typical for a symbiotic outburst.  The system again showed
brightening in early 1998 as noticed by Wakuda.  Wakuda's survey on his
patrol films revealed another brightening in 1984, which is also included
in Table 1 and Figure 1.

% figure 1
\begin{figure}
  \begin{center}
  \includegraphics[angle=0,height=6.5cm]{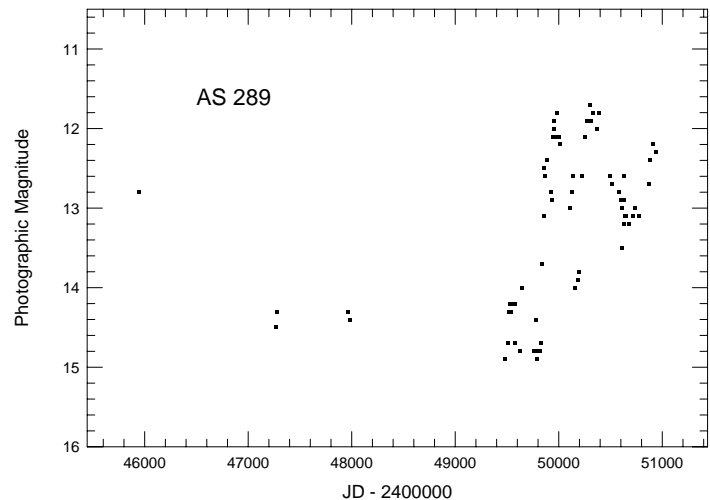}
  \caption{Overall light curve of AS 289} \label{fig-1}
  \end{center}
\end{figure}

\section{Discussion}
   The present observation is the first that revealed the significant
variation of the symbiotic star AS 289.  The variation is globally
characterized by episodic outbursts observed in 1984 and 1995.  This
interval of outbursts is typical for those of classical symbiotic variables
(the limited observations between these two outbursts may have missed some
other brightenings, though).
However, the early 1996 fading episode is unique, and remarkably similar
to eclipses seen in FG Ser = AS 296 (Munari et al. 1992) and V1413 Aql = 
AS 338 (Wakuda 1988; Munari 1992).  Figure 2 shows the detailed light curve
of the current outburst.  We suspect the eclipse is responsible for
the fading episode of AS 289 during outburst.  The profile of light curve
suggests the eclipse may be grazing total, while the brightness at minimum
is considerably brighter than the pre-outburst level, implying that
a certain fraction of the outbursting source remain uneclipsed.
It is not still certain regarding the periodicity, but considering the duration
of the fading ($\sim$ 100 d) comparable to that of FG Ser (Munari et al.
1992), the orbital period may be an order of $\sim$ 600-700 d.  This period
may imply that the relatively faint level in late 1997 may be somehow
related to eclipse, or that the next eclipse was unfortunately missed
during the solar conjunction period.

% figure 2
\begin{figure}
  \begin{center}
  \includegraphics[angle=0,height=6.5cm]{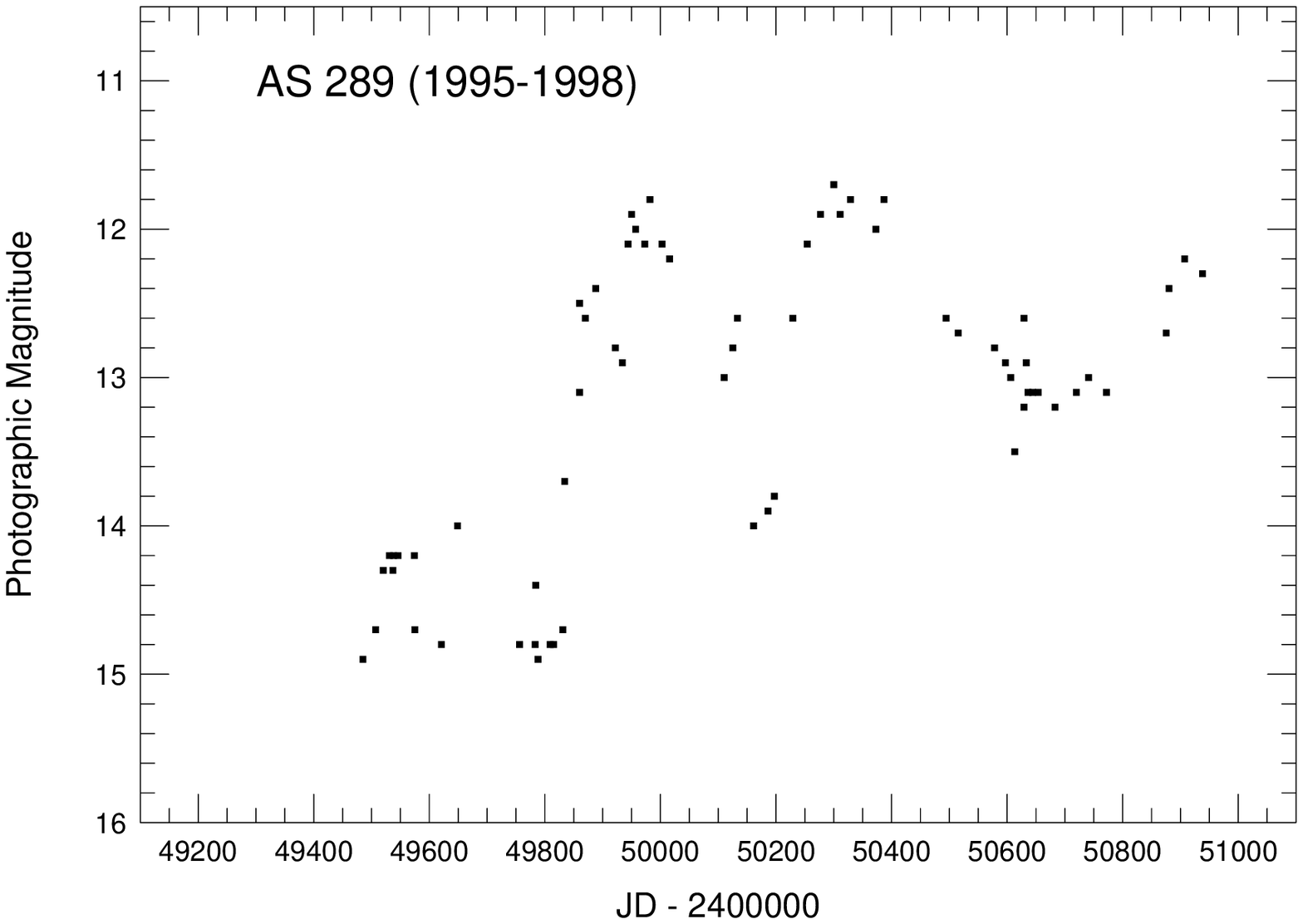}
  \caption{Light curve of AS 289 in 1995--1998} \label{fig-2}
  \end{center}
\end{figure}

\end{document}